\documentclass[twocolumn,prl,aps,showpacs,superscriptaddress]{revtex4}
\usepackage{graphicx}
\usepackage{dcolumn}
\usepackage{bm}

\begin{document}
\bibliographystyle{prsty1}
\title{Asymmetric Magnetization Reversal in a Single Exchange-Biased Micro Bar}
\author{T. Gredig}
\email{tgredig@csulb.edu} 
\affiliation{Department of  Physics and Astronomy,California State University Long Beach, 1250 Bellflower Blvd., Long Beach, CA 90840}
\affiliation{School of Physics and Astronomy, University of Minnesota, 116 Church St. SE, Minneapolis, MN 55455}

\author{M. Tondra}
\affiliation{Diagnostic Biosensors, 1712 Brook Ave. SE, Minneapolis, MN 55414}
\date{\today}

\begin{abstract}
The asymmetric magnetization reversal is studied in a single exchange-biased microbar of 1.5~$\mu$m $\times$ 13~$\mu$m with anisotropic magnetoresistance and magnetic force microscopy. The particle has a moment of less than 10$^{-9}$ emu and is not accessible with standard magnetometry. The asymmetric hysteresis loop of CoFe/CrMnPt shows a repeatable rotation process, followed by an irreversible nucleation process that is marked by jumps in the magnetoresistance. The induced unidirectional anisotropy enhances the rotation process in one branch of the hysteresis loop, followed by a sped up nucleation process. Imprinted ferromagnetic domain patterns left behind by the antiferromagnet are observed after the nucleation process occurred but before complete saturation is reached.
\end{abstract}


\maketitle

\section{Introduction}

The asymmetries of the magnetic reversal mechanism in exchange-biased thin films have attracted interest recently.\cite{camarero_origin_2005,blomqvist:107203} One outstanding problem is to understand the reversal mechanism of small particles, which have a size comparable to the characteristic length of their ferromagnetic domains. Several domain models have been put forward to explain the coupling mechanism between ferromagnetic and antiferromagnetic materials, but details regarding the important reversal mechanism remain unresolved.\cite{Stiles_1999,keller:014431} 

Magnetic reversal is usually characterized with hysteresis loops, or MOKE measurements that can measure both the perpendicular and parallel magnetic component in a thin film.\cite{tillmanns:202512} In exchange-bias thin films, asymmetries of the reversal have been observed and discussed previously.\cite{tsang:2605,ambrose:7222,PhysRevLett.84.3986, PhysRevLett.91.187202} The unidirectional anisotropy induced by the antiferromagnet allows for a complex magnetization reversal. Asymmetric reversal in exchange biased films can originate either from a misalignment\cite{beckmann_asymmetric_2003,camarero_origin_2005} or local fluctuations of the anisotropy axis.\cite{czapkiewicz_thermally_2008} Indeed, a sample can exhibit two separate reversal behaviors depending on the directionality of the applied field.\cite{Fitzsimmons_2000} One reversal could be dominated by a rotation mechanism, and another mechanism is triggered through a nucleation process, where a domain can rotate quickly through almost 180 degrees and then grows by propagation of domain walls. The interplay of the two mechanisms is explored in exchange-biased CoFe/CrMnPt samples.

Here, a rectangular magnetic bar of dimensions 1.5 $\mu$m $\times$ 13 $\mu$m is investigated with anisotropic magnetoresistance (AMR) and magnetic force microscopy (MFM) with a variable in-plane magnetic field. While other studies on micro- and nano-patterned exchange-biased systems focus on the collective behavior of arrays,\cite{sort:067201,bollero:022508,li:072501,temst:10K117,jung:10K113,pokhil:6887} here the focus is on a single element. Such an element has dimensions achievable by calculations using micromagnetic simulations.\cite{saha:073901} The magnetoresistance measurements exemplify the asymmetry in the reversal, whereas the magnetic force microscopy images show the distribution of the magnetic domains at different stages of the reversal. Unlike vibrating sample magnetometry (VSM) or SQUID measurements, the AMR measurements can be performed on particles with magnetic moments under $10^{-9}$ emu. The combination of AMR with MFM allows to compare the hysteresis loop and micromagnetic structure of a small multi-domain element.

\begin{figure}
\includegraphics[width=8cm]{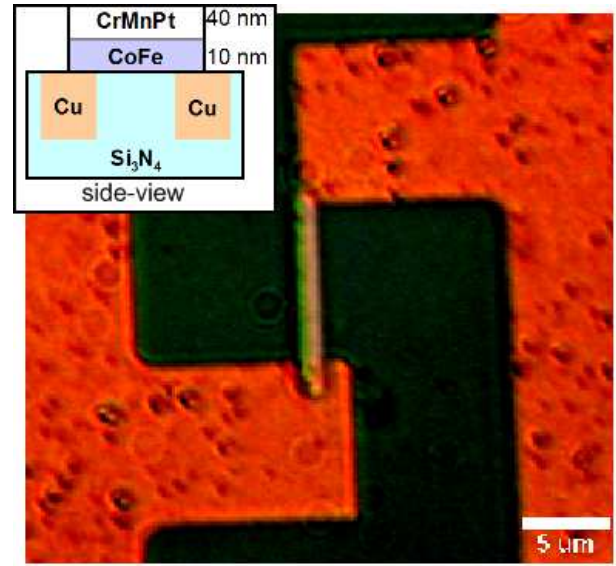} 
\caption{\label{fig:particle} Optical micrograph of exchange biased CoFe/CrMnPt microbar connected to two Cu terminals. The inset is a schematic side view of the embedded Cu leads and the exchange-biased bilayer.}
\end{figure}

\section{Experiment}

Substrates were prepared lithographically such that copper lines could be embedded into a non-conductive Si$_3$N$_4$ matrix. After an electro-chemical polish, the surface was smooth enough to deposit the F/AF bilayer. The exchange bias bilayer of Co$_{50}$Fe$_{50}$ and Cr$_{45.5}$Mn$_{45.5}$Pt$_9$ was sputter deposited and  photolithography placed on top of two copper elements for resistance measurements. The approach of embedded copper lines ensures that the ferromagnetic microbar is not bent at each of its endpoints, see Fig.~\ref{fig:particle}. The nominal thickness of CrMnPt is determined to be 40 nm, slightly larger than the critical thickness of 25 nm necessary for exchange bias in CrMnPt.\cite{soeya_enhanced_1997} The antiferromagnet contains equal portions of Mn and Cr and 9\% of Pt in order to maximize the blocking temperature. For this material, the blocking temperature has been measured to be near 590 K,\cite{soeya_enhanced_1997,hoshiya_spin-valve_1997,soeya_exchange_2002} thus the bilayer exhibits exchange bias at room temperature. Two samples were prepared that only varied in the thickness of the ferromagnet. In the first sample, the CoFe layer is 10 nm, and the second sample, the ferromagnet is 15 nm thick.

A magnetic field was applied during an annealing process to set the exchange bias field along the long-axis of the bar. Once set, the sample exhibits a reproducible unidirectional anisotropy at room temperature.

\begin{figure}
\includegraphics[width=8cm]{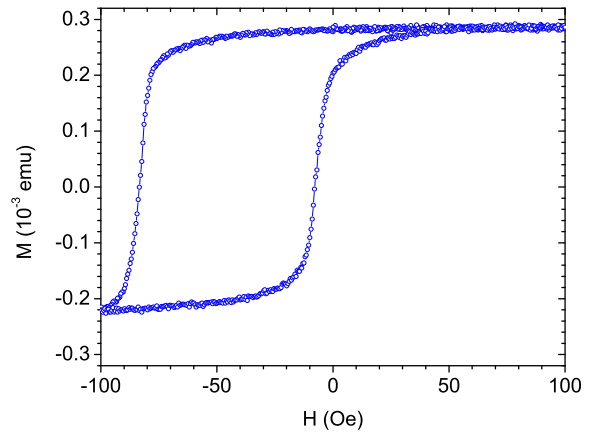} 
\caption{\label{fig:VSMHystLoop636-12L} Easy axis hysteresis loop of a CoFe(15nm)/CrMnPt(40nm) bilayer exhibits typical exchange bias behavior at room temperature. The measurement was performed with a vibrating sample magnetometer.}
\end{figure}

A witness sample was deposited as a thin film that could be measured in a vibrating sample magnetometer (VSM) for comparison with a single exchange-biased bar. In this case, the hysteresis loop for the easy axis is completely shifted; i.e. the exchange bias of 48 Oe is larger than the coercivity of 38 Oe, see Fig.~\ref{fig:VSMHystLoop636-12L}. As expected, these values are smaller by about 20\% than those found in the micro-sized elements using MFM and AMR. The hard axis hysteresis loop of the thin film has almost no coercivity and saturates near 150 Oe. 

The estimated magnetic moment of the CoFe bar is only $4\cdot 10^{-10}$~emu, much lower than the detectable limit by either a SQUID or VSM device. Since the resistance scales with the geometry of the device, AMR is able to provide detailed data about the magnetization reversal, even though it does not measure the absolute magnetic moment.

The sample is mounted on a rotating stage of a modified magnetic force microscope that allows the application of external magnetic fields. An electro-magnet with coils that can provide up to 1200 Oe in-plane magnetic field at its center surrounds the MFM system. The magnetic field is sufficient to saturate the sample at room temperature. The sample is placed inside the magnetic field with standard pseudo four-terminal resistance apparatus connected. The magnetization reversal and domain configuration in these particles was measured by magnetic force microscopy using a standard CoCr coated Si cantilever.

\section{Results and Discussion}

Applying a saturating magnetic field of 1100 Oe, the sample is rotated, and the resistance is measured as a function of angle. The response is consistent for a polycrystalline magnetic sample and shows the near typical $\cos^2 \theta$ behavior, where $\theta$ is the angle between the current and the magnetization. In the 10nm thick CoFe sample, the magnetoresistance amplitude $\Delta \rho$ is 1.503 $\Omega$, which corresponds to 0.76\% AMR. As expected, the AMR value in thin films is smaller compared to bulk materials due to additional surface scattering.

\begin{figure}
\includegraphics[width=8cm]{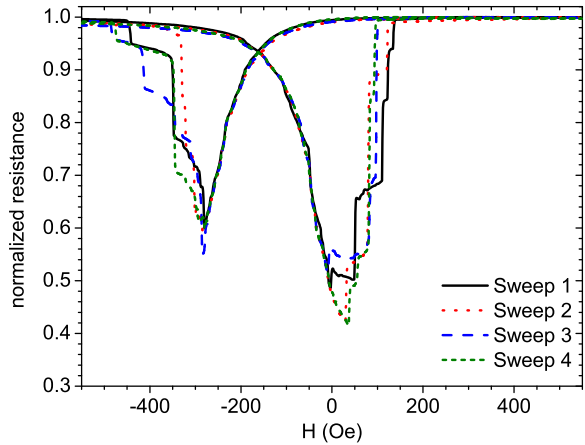} 
\caption{\label{fig:3Loops} Hysteresis loops as measured with anisotropic magnetoresistance for three different runs of a single exchange-biased CoFe(10nm)/CrMnPt bilayer. The magnetization reversal is asymmetric due to the exchange bias. A repeatable smooth rotation is followed by an irreversible nucleation process, which results in steps in the resistance measurement. The resistance is normalized to the full AMR amplitude, where a value of 1.0 (0.0) indicates the magnetization is parallel (perpendicular) with the current.}
\end{figure}

The orientation is set by monitoring the resistance as the angle between the current and the field is changed. The precise angle of the current direction is determined to align the sample. For the magnetic reversal measurements, several consecutive curves are measured as the magnetic field is swept from +1000 Oe to -1000 Oe, see Fig.~\ref{fig:3Loops}. As clearly shown, the hysteresis loop is shifted and exhibits an asymmetry. It should be noted that the asymmetry observed in the CrMnPt system is more subtle and differs from the well-studied Co/CoO system.\cite{welp:7726,hoffmann:097203,gredig_temperature_2006}. In the later case, the strong asymmetry is due to a reconfiguration of the AF lattice during the first magnetization reversal. Here, a weakened asymmetry persists unrelated to the training effect.\cite{binek:067201}

The left magnetization reversal branch is noticeably smaller than the right reversal branch. The asymmetry is not due to the scale or geometry of the bar, as it is also observed in the thin film (Fig.~\ref{fig:VSMHystLoop636-12L}) and eleswhere.\cite{gredig_temperature_2006} In essence, the left reversal branch has fewer magnetic domains that are perpendicular to the current direction. For the right branch, more domain rotation is observed. It is noted that during the rotation, a smooth reproducible curve is measured, and once the curve peaks, several steps in the resistance are observed (Fig.~\ref{fig:3Loops}). These steps are associated with growth of reversed domains and a nucleation process that occurs after the domain rotation is limited. The nucleation process is not reproducible and probably depends on the rate of magnetic field change and its exact direction. The width over which the nucleation process occurs is wider in the left branch (200 Oe) as compared to the right branch (100 Oe).

\begin{figure}
\includegraphics[width=8cm]{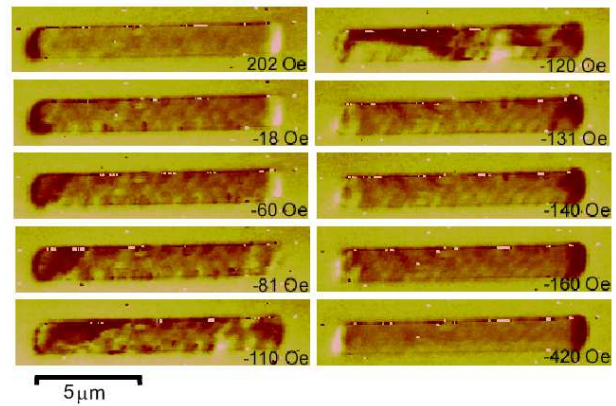} 
\caption{\label{fig:mag_rev1} Magnetic force microscopy image of the left magnetization reversal branch for an exchange biased CoFe(15nm)/CrMnPt element. The induced exchange bias field delays the reversal of the magnetization, which occurs near -130 Oe. }
\end{figure}

The small magnetization of the 10 nm CoFe covered with 40 nm CrMnPt AF layer was insufficient for a reliable magnetic response from the MFM. Thus, a second sample fabricated with the same procedural steps, but with a 15 nm CoFe layer is used to analyze the reversal process with MFM. In Fig.~\ref{fig:mag_rev1}, the bar is fully magnetized at 200 Oe, as the field is reduced, magnetic domains start to form near -20 Oe. The domains begin to rotate until about -80 Oe, at which point, there is propagation of the domain along the length of the bar. The propagation is almost completed near -130 Oe, at which point, only small enclaves near the edge are left before the sample reaches complete saturation.

\begin{figure}
\includegraphics[width=8cm]{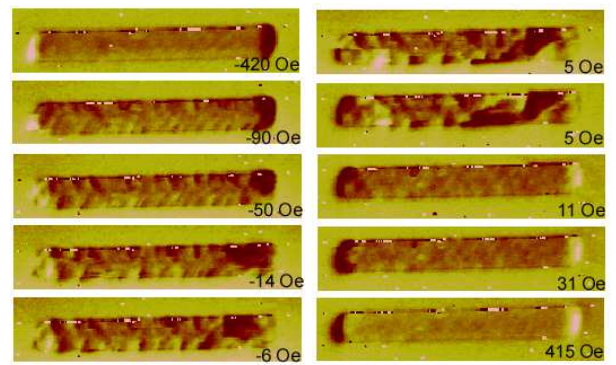} 
\caption{\label{fig:mag_rev2} Magnetic force microscopy image of the right magnetization reversal branch for an exchange biased CoFe(15nm)/CrMnPt element. The induced exchange bias field exerts a torque that leads to a low-field reversal of the magnetization at +5 Oe. }
\end{figure}

The magnetization reversal in the opposite direction is qualitatively similar, but differs in a few key features. As seen in Fig.~\ref{fig:mag_rev2}, in a field of -420~Oe, two poles at the edge indicate the the bar is fully magnetized. Even before removing the field completely, domains start to rotate and form at -90 Oe, because the antiferromagnetic pinning energy favors a reversal of the domains. A clear pattern of the rotated domains starts to emerge and intensify as the field is reduced to -5 Oe. Near +5 Oe, nucleation and propagation of magnetic domains starts to occur, such that the reversal is almost completed at +30 Oe. However, some "imprinted" domains are still visible and much larger fields (+400 Oe) are necessary to wipe out those marks.

During the second reversal, the domain rotation is more pronounced. At -6 Oe, there are many small ripples in the magnetic image that indicate rotated domains. This is in agreement with the AMR measurements, where this branch shows a deeper resistance minimum, which would indicate that more magnetic domains are perpendicular to the current directions than in the previous branch. 

The competition of the unidirectional anisotropy with the externally applied magnetic field leads to less rotation in larger external fields, since the antiferromagnet prefers perpendicular coupling that leads to rotation. If the applied field is small, the antiferromagnet causes rotation of domains, as seen in the right branch. To the contrary, in the left branch the ferromagnet is pinned and only large applied fields will reverse it, in which case the nucleation process is more important. Interestingly, even though the magnetization reversal is completed as judged by the clear poles at either end, at +11 Oe of the right branch (Fig.~\ref{fig:mag_rev2}), small ripples persist. A clear demonstration that the domain pattern is not fully erased yet.\cite{welp:7726} These ripples indicate a connection to the AF structure that is inherent to the thin film after the in-field cooling procedure. At  high fields (+415 Oe), the film is completely saturated and most ripples have disappeared.

The domain size measured in the left and right branch of the reversal appear of similar size unlike what was observed in Co/CoO thin films. In the latter case, a distinct difference in domain sizes was inferred from polarized neutron diffraction data.\cite{Gierlings2005}

\section{Conclusions}

The use of anisotropic magnetoresistance allows the study of the asymmetric magnetization reversal of individual magnetic particles with moments less than $10^{-9}$~emu. The field sweep data of a single particle coupled with magnetic force microscopy results in a microscopic understanding of the reversal mechanism in exchange-biased bilayers. The investigation of CoFe/CrMnPt bilayers showed an asymmetric reversal that has two parts. The first reversal process, a rotation of domains, is well reproducible and originates from the AF/F coupling energy that prefers domains perpendicular to the applied field. In the right branch, more rotation is observed as the external field to reverse all domains is smaller. The second process, a nucleation process, differs for each sweep with distinct resistance jumps. Small ripples are left behind after the reversal due to a specific antiferromagnetic lock-in pattern created during the annealing process.

\begin{acknowledgments}
The authors would like to thank E. Dan Dahlberg for helpful discussions and comments to the manuscript. Samples were prepared with the help of NVE Corporation. This research was supported primarily by the MRSEC Program of the National Science Foundation under Award Number DMR-9809364 and DMR-0212302.
\end{acknowledgments}

\end{document}